\pdfoutput=1

\documentclass[lettersize,journal]{IEEEtran}

\usepackage{amsmath,amsfonts,bm}

\usepackage{graphicx}
\usepackage{array, booktabs, multirow, makecell, colortbl}

\usepackage[caption=false,font=footnotesize]{subfig}

\usepackage{stfloats}
\usepackage[section]{placeins}
\usepackage{float}

\usepackage{algorithm}
\usepackage{algpseudocode}

\usepackage{listings}
\usepackage{enumerate}

\usepackage{cite}
\usepackage{textcomp}
\usepackage{xcolor}
\usepackage{url}

\usepackage{etoolbox}
\usepackage{verbatim}
\usepackage{setspace}

\usepackage{titlesec}   
\titlespacing\section{0pt}{6pt plus 4pt minus 2pt}{0pt plus 2pt minus 2pt}  
\titlespacing\subsection{0pt}{6pt plus 4pt minus 2pt}{0pt plus 2pt minus 2pt}  
\IEEEaftertitletext{\vspace{-2\baselineskip}} 

\begin{document}

\title {Subspace Fitting Approach for Wideband Near-Field Localization}

\author{Ruiyun Zhang, Zhaolin Wang, Zhiqing Wei, Yuanwei Liu,~\IEEEmembership{Fellow,~IEEE,} Zehui Xiong, Zhiyong Feng

\thanks{Ruiyun Zhang, Zhiqing Wei, and Zhiyong Feng are with the School of Information and Communication Engineering, Beijing University of Posts and Telecommunications, China (e-mail: zhangruiyun@bupt.edu.cn; weizhiqing@bupt.edu.cn; fengzy@bupt.edu.cn).}
\thanks{Yuanwei Liu, Zhaolin Wang are with the Department of Electrical and Electronic Engineering, The University of Hong Kong, Hong Kong (e-mail: yuanwei@hku.hk; zhaolin.wang@hku.hk).}
\thanks{Zehui Xiong is with the School of Electronics, Electrical Engineering and Computer Science, Queen's University Belfast, United Kingdom
(e-mail: z.xiong@qub.ac.uk).}}

%

\maketitle

\begin{abstract}
Two subspace fitting approaches are proposed for wideband near-field localization. Unlike in conventional far-field systems, where distance and angle can be estimated separately, spherical wave propagation in near-field systems couples these parameters. We therefore derive a frequency-domain near-field signal model for multi-target wideband systems and develop a subspace fitting-based MUSIC method that jointly estimates distance and angle. To reduce complexity, a Fresnel approximation MUSIC algorithm is further introduced to decouple the distance and angle parameters. Numerical results verify the effectiveness of both proposed approaches.
\end{abstract}

\begin{IEEEkeywords}
MUSIC, Subspace fitting, Wideband near-field localization.
\end{IEEEkeywords}

\IEEEpeerreviewmaketitle

\vspace{-3mm}
\section{Introduction}
\IEEEPARstart{A}s a fundamental technology in modern wireless systems, multiple-input multiple-output (MIMO) architectures enable both communication and sensing capabilities simultaneously~\cite{bjornson2023twenty}. The resemblance in structure and signal processing between MIMO-based communication and sensing systems positions MIMO technology as a crucial enabler of integrated sensing and communication (ISAC), a capability that is now deemed essential for future wireless infrastructure~\cite{liu2022integrated}.


Nowadays, the evolution of the sixth-generation (6G) wireless networks is driving a paradigm shift toward extra-large MIMO (XL-MIMO) configurations~\cite{wang2024tutorial}. Next-generation XL-MIMO systems are designed to incorporate antenna counts ranging from hundreds to thousands of discrete elements, representing a significant scaling beyond conventional arrays. This quantum leap in array dimensionality precipitates two transformative effects~\cite{wang2025performance}: (1) an order-of-magnitude increase in effective aperture size, and (2) a complete reconfiguration of electromagnetic wave behavior. Among the most significant ramifications is the orders-of-magnitude expansion of the reactive near-field boundary around transmission arrays. For a 1-meter aperture array operating at 30 GHz, the near-field region extends to approximately 200 meters. Consequently, most targets are expected to be located within this near-field domain. Electrodynamic analysis reveals that such configurations exhibit Fresnel-zone wavefront curvature, presenting a stark departure from the traditional far-field Fraunhofer approximation. Consequently, the entire theoretical framework for assessing both data transmission and environmental sensing capabilities requires comprehensive re-engineering to account for these near-field propagation conditions, which enables near-field sensing (NISE) to achieve distance estimation without requiring the wideband typically needed in far-field scenarios.

Conventional wideband far-field sensing relies on the plane wave assumption, limiting its capability to estimate only the angular parameters of signal sources, while distance resolution is fundamentally constrained by system bandwidth~\cite{wang2017unified}. As a result, distance and angle estimation can be performed independently in the spatial and frequency domains of conventional wideband far-field systems. Prior works~\cite{10048770} estimated decoupled angle and distance for ISAC downlink and uplink, respectively.

However, the spherical wavefronts in near-field regimes generate high coupling between distance and angle information across signal domains, and wideband NISE utilizes spherical wave propagation characteristics and received signal wavefront curvature to enable high-precision joint estimation of both azimuth angle and radial distance~\cite{9693928}. Wang et.al revealed how near-field effects enhance wideband ISAC capabilities and proposed new design paradigms~\cite{wang2024rethinking}, and derived Cramer-Rao bounds for NISE in wideband orthogonal frequency division multiplexing (OFDM) systems~\cite{wang2025performance}. Luo et.al transformed beam squint from a impairment to an advantage for ISAC, enabling simultaneous communication and localization with reduced overhead~\cite{10271123}. Although existing studies have demonstrated the wideband capabilities and strong angle-distance coupling characteristics of near-field systems, there remains a significant gap in developing high-precision parameter estimation algorithms capable of supporting super-resolution angle and distance estimation for emerging ISAC applications.

Motivated by the above, this paper investigates wideband near-field localization using two subspace fitting-based multiple signal classification (MUSIC) approaches. The main contributions of this paper are summarized below.

\begin{itemize}
\item We derive a frequency-domain signal model for multi-target OFDM systems operating in the near-field region. Based on this model, we develop a subspace fitting-based MUSIC method to simultaneously estimate the distance and angle parameters of multiple targets.

\item To further enhance computational efficiency, we introduce a low-complexity MUSIC algorithm that leverages the Fresnel approximation, significantly reducing the search space dimensionality of the proposed algorithm.

\item Finally, numerical results are provided to validate the effectiveness of the proposed designs.
\end{itemize}

\vspace{-0.5mm}
\section{Wideband Near-Field Signal Model}
\label{sec:sig}
\begin{figure}[tp]
\begin{center}
\includegraphics[width=0.60\linewidth]{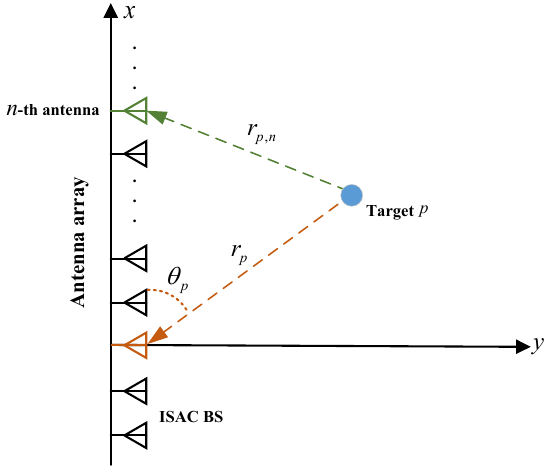}
\end{center}

\caption{Illustration of the considered near-field system model.}
\label{fig:fig-1}
\vspace{-4mm}\end{figure}

This paper investigates a near-field communication system, comprising a base station (BS) equipped with an $N$-element uniform linear array (ULA) and $P$ single-antenna targets. The OFDM technique is adopted to effective exploit the large bandwidth. As depicted in Fig.~\ref{fig:fig-1}, the Cartesian coordinate system is established with its origin at the center of the ULA, aligned along the $x$-axis. The antenna elements are spaced at $d = {\lambda _c} / 2$, where $\lambda _c$ denotes the center carrier wavelength.

The spatial position of the $n$-th antenna element at the BS is defined by the coordinates $\left( {{\delta _n}d,0} \right)$, where ${\delta _n} = n - 1 - (N - 1)/2, n = 1, ..., N$. The BS receives wideband OFDM signals emitted or scattererd by $P$ targets located within the near-field region, where the target-BS distances satisfy the Rayleigh criterion $2{D^2}/{\lambda _c}$. Here, $D = N d$ represents the aperture of the antenna array. Under the quasi-static channel assumption, all targets are considered stationary or exhibiting sufficiently low mobility to render Doppler effects negligible. The position of the $p$-th target is characterized by its radial distance $r_p$ from the Cartesian coordinate origin and angular position $\theta_p$ relative to the $x$-axis. The spatial locations of the $p$-th target are consequently defined as $(r_p\cos \theta_p ,r_p\sin \theta_p)$. The Euclidean distance between the $n$-th array element and the $p$-th target, denoted as $r_{p,n}$, is expressed as
\begin{equation}
\label{eq:eq-1}
{r_{p,n}} = \sqrt {{r_p^2} + \delta _n^2{d^2} - 2r_p{\delta _n}d\cos \theta_p }.
\vspace{-1mm}\end{equation}

Let $\tau _{p,n} = \frac{r_{p,n}}{c}$ represent the propagation delay, where $c$ denotes the speed of light. An OFDM frame consisting of $K$ OFDM symbols is considered, with each symbol having an overall signal bandwidth of $M\Delta f = B$ and a total symbol duration of ${T_t} = {T_d} + {T_g}$. Here, $M$ denotes the number of subcarriers, and $\Delta f = 1/{T_d}$ represents the subcarrier spacing. Additionally, ${T_g}$ and ${T_d}$ denote the cyclic prefix (CP) duration and the elementary OFDM symbol duration, respectively. The complex baseband signal transmitted by the $p$-th target over $K$ symbols (including CP) can be expressed as follows:
\begin{equation}
\label{eq:eq-2}
{{x}_p}(t) \!=\! \dfrac{1}{{\sqrt {{M}} }} \sum\limits_{k = 0}^{{K} - 1} \sum\limits_{m = 0}^{{M} - 1}  {{{s}}_{k,m,p}}{e^{j2\pi m\Delta f t }} \\
{\rm{rect}} \left( \frac{{t - k{T_t}}}{{{T_t}}} \right),
\end{equation}
where ${{{s}}_{k,m,p}}$ denotes the complex-valued data symbol transmitted by the $p$-th target on the $m$-th subcarrier of the $k$-th OFDM symbol. ${\rm{rect}}( \cdot )$ denotes the rectangular window function, which defines the temporal structure of each OFDM symbol.

Then, the signal received by the BS at the $n$-th antenna, transmitted by the $p$-th target, can be modeled as ${y_{p,n}}(t) = {{\beta _{p,n}}} {x_p}(t - \frac{{{r_{p,n}}}}{c}){e^{ - j2\pi {f_c}\frac{{{r_{p,n}}}}{c}}}$, where ${\beta _{p,n}}$ denotes the complex channel coefficient incorporating path loss and phase shift, and $f_c$ represents the carrier frequency. In the uplink transmission scenario, the received signal at the $n$-th array element comprises the superposition of signals from all $P$ targets, which can be mathematically expressed as follows:
\begin{equation}
\label{eq:eq-3}
{y_{n}}(t) = \sum\limits_{p = 1}^{P} {{\beta _{p,n}}} {x_p}(t - \frac{{{r_{p,n}}}}{c}){e^{ - j2\pi {f_c}\frac{{{r_{p,n}}}}{c}}} + {w_{n}}(t),\\
\end{equation}
where ${w_{n}}(t)$ denotes additive stationary Gaussian noise (AWGN) with zero-mean and variance $\alpha_n^2$. Through discrete-time signal processing, the received OFDM signal can be sampled at time instants ${y_n}(t)$ at $t = k{T_t} + i{T_s}$ after cyclic prefix removal, where $T_s = 1/B = 1/(M \Delta f)$ denotes the sampling interval. The resulting discrete-time signal model is
\begin{equation}
\label{eq:eq-4}
\begin{array}{c}
\begin{aligned}
&{y_n}(k{T_t} + i{T_s}) \buildrel \Delta \over = {y_{k,n}}\left[ i \right]\\
\!=& \frac{1}{{\sqrt M }}\!\sum\limits_{p = 1}^{P} {\!\sum\limits_{m = 0}^{M - 1} {{\beta _{p,n,m}}} {s_{k,m,p}}{e^{j2\pi m\Delta f\left( {i{T_s} - {\tau _{p,n}}} \right)}}{e^{ - j2\pi {f_c}{\tau _{p,n}}}}}\\
\!=& \frac{1}{{\sqrt M }}\!\sum\limits_{p = 1}^{P} {\!\sum\limits_{m = 0}^{M - 1} {{\!\beta _{p,n,m}}} {s_{k,m,p}}{e^{j2\pi \frac{{mi}}{M}}}{e^{ - j2\pi {f_m}{\tau _{p,n}}}}},
\end{aligned}
\end{array}\
\vspace{-1mm}\end{equation}where $f_m = f_c + m \Delta f$ denotes the frequency of the $m$-th subcarrier. The frequency-dependent channel attenuation coefficient is calculated as ${\beta _{p,n,m}} = \frac{c}{4 \pi f_m r_{p,n}}$~\cite{10271123}. Given that the target-array distance significantly exceeds the array aperture (i.e., $r_{p,n} \gg D$), we can reasonably approximate the channel coefficients across the array elements as ${\beta _{p,1,m}} \approx  \ldots  \approx {\beta _{p,N,m}} \approx {\beta _{p,m}} = \frac{c}{{4\pi {f_m}{r_p}}}$. Furthermore, under the practical assumption that the carrier frequency dominates the bandwidth (${f_c} \gg M\Delta f$), the frequency-dependent attenuation can be simplified to a frequency-independent model: ${\beta _{p,0}} \approx  \ldots  \approx {\beta _{p, M - 1}} \approx {\beta _{p}} = \frac{c}{{4\pi {f_c}{r_p}}}$.
\vspace{-1mm}
The received signal vector at the BS, ${{\bf{y}}_k}\left[ i \right] = {\left[ {\begin{array}{*{20}{c}}
{{y_{k,1}}\left[ i \right],}&{{y_{k,2}}\left[ i \right],}&{ \ldots ,}&{{y_{k,N}}\left[ i \right]}
\end{array}} \right]^T}$, comprising signals from all $N$ array elements, is calculated as
\begin{equation}
\label{eq:eq-5}
\begin{array}{c}
\begin{aligned}
{{\bf{y}}_k}\left[ i \right] = \frac{1}{{\sqrt M }}\sum\limits_{m = 0}^{M - 1} { {\bm \beta}} \odot {{\bf{A}}_m}\left( \bm{{r, \theta }} \right){{\bf{s}}_{k,m}}{e^{j2\pi \frac{{mi}}{M}}},
\end{aligned}
\end{array}\
\end{equation}where ${ {\bm \beta}} = \left[ {\begin{array}{*{20}{c}}{{\beta _{1}},}&{ \ldots ,}&{{\beta _{P}}}
\end{array}} \right]$ denotes the path loss vector, and the near-field steering matrix ${{\bf{A}}_m}\left( \bm {r,\theta } \right)$ is defined as $[ {\begin{array}{*{20}{c}}
{{{\bf{a}}_m}({r_1},{\theta _1}),}&{ \ldots ,}&{{{\bf{a}}_m}({r_P},{\theta _P})}
\end{array}}]$, where ${{\bf{a}}_m}({r_p},{\theta _p}) = {[ {\begin{array}{*{20}{c}}{{e^{ - j2\pi {f_m}{\tau _{p,1}}}},}&{ \ldots ,}&{{e^{ - j2\pi {f_m}{\tau _{p,N}}}}}\end{array}} ]^T} \in {{\mathbb C}^{N \times 1}}$ denotes the near-field steering vector with respect to $f_m$ and $r_{p,n}$. Additionally, ${{\bf{s}}_{k,m}} = {\left[ {\begin{array}{*{20}{c}}
{{s_{k,m,1}},}&{ \ldots ,}&{{s_{k,m,P}}}
\end{array}} \right]^T}$ denotes the transmitted symbol vector for the $m$-th subcarrier in the $k$-th OFDM symbol.

Through the discrete Fourier transform (DFT) processing of the received OFDM sequence, the frequency-domain representation ${{\bf{y}}_{k,m}}$ is given by
\begin{equation}
\label{eq:eq-6}
\begin{array}{c}
\begin{aligned}
{{\bf{y}}_{k,m}} \!\!=\! {\rm DFT}{\left( {{{\bf{y}}_k}\!\left[ i \right]} \right)_m} \!\!+\! {{\bf{w}}_{k,m}} \!\!=\! { {\bm \beta} } \! \odot\! {{\bf{A}}_m}\!\left( \bm {r,\theta } \right) {{\bf{s}}_{k, m}} \!\!+\! {{\bf{w}}_{k,m}},
\end{aligned}
\end{array}\
\end{equation}
where ${{\bf{w}}_{k,m}}$ denotes the frequency-domain noise vector. The received signal matrix  ${{\bf{Y}}_m} = \left[ {\begin{array}{*{20}{c}}
{{{\bf{y}}_{1,m}},}&{ \ldots ,}&{{{\bf{y}}_{K,m}}}
\end{array}} \right] \in {{\mathbb C}^{N \times K}}$ for the $m$-th subcarrier across $K$ OFDM symbols is
\begin{equation}
\label{eq:eq-7}
\begin{array}{c}
\begin{aligned}
{{\bf{Y}}_m} &= { {\bm \beta} } \odot {{\bf{A}}_m}\left( \bm {r,\theta } \right){{\bf{S}}_m} + {{\bf{W}}_m},\\
\end{aligned}
\end{array}\
\end{equation}where ${{\bf{S}}_m} = \left[ {\begin{array}{*{20}{c}}
{{{\bf{s}}_{0, m}},}&{ \ldots ,}&{{{\bf{s}}_{K - 1, m}}}
\end{array}} \right] \in {{\mathbb C}^{P \times K}}$ denotes the data symbol matrix on the $m$-th subcarrier across all $K$ symbols, and ${{\bf{W}}_m} = \left[ {\begin{array}{*{20}{c}} {{{\bf{w}}_{1,m}},}&{ \ldots ,}&{{{\bf{w}}_{K,m}}} \end{array}} \right] \in {{\mathbb C}^{N \times K}}$ represents the frequency-domain noise matrix.

\section{Joint Distance-Angle Estimation}
\label{sec:est}

\subsection{Subspace Fitting-based MUSIC Method}

For each received subcarrier ${{\bf{Y}}_m}$ in Eq.~(\ref{eq:eq-7}), the 2D-MUSIC algorithm can be employed for joint distance-angle estimation. Specifically, the covariance matrix of the received noisy symbols, calculated from $K$ snapshots, is given by\vspace{1mm}
\begin{equation}
\label{eq:eq-8}
\begin{array}{c}
\begin{aligned}
{{\bf{\Sigma}}_m} = \frac{1}{K}{{\bf{Y}}_m}{\bf{Y}}_m^H \in {{\mathbb C}^{N \times N}}.
\end{aligned}
\end{array}\
\end{equation}
Through eigenvalue decomposition of the covariance matrix ${{\bf{\Sigma}}_m}$, we obtain\vspace{-1mm}
\begin{equation}
\label{eq:eq-9}
\begin{array}{c}
\begin{aligned}
{{\bf{\Sigma}}_m} = {\bm{\mathcal{V}}_{sm}}{{\bf{\Lambda }}_{sm}}{\bm{\mathcal{V}}}_{sm}^H + {{\bm{\mathcal{U}}}_{nm}}{{\bf{\Lambda }}_{nm}}{\bm{\mathcal{U}}}_{nm}^H,
\end{aligned}
\end{array}\
\end{equation}
where ${{\bm{\mathcal{V}}}_{sm}} = \left[ {{{\bf{v}}_{m1}}, \ldots ,{{\bf{v}}_{{mP}}}} \right]$ denotes the signal subspace matrix, and ${{\bf{\Lambda }}_{sm}} = \mathrm{diag} \left\{ {{\lambda _{m1}}, \ldots ,{\lambda _{{mP}}}} \right\}$ contains the corresponding eigenvalues in descending order. Additionally, the noise subspace is characteristic by ${{\bm{\mathcal{U}}}_{nm}} = \left[ {{{\bf{u}}_{m({P + 1})}}, \ldots ,{{\bf{u}}_{mN}}} \right]$ with associated eigenvalues ${{\bf{\Lambda }}_{nm}} = \mathrm{diag} \left\{ {{\lambda _{m{({P} + 1)}}}, \ldots ,{\lambda _{mN}}} \right\}$.

In this work, the subspace fitting-based MUSIC algorithm~\cite{10934790} is employed for joint distance-angle estimation across all subcarriers. For the $m$-th subcarrier, the MUSIC optimization problem is formulated through the following objective function:
\begin{equation}
\label{eq:eq-11}
\begin{array}{c}
\begin{aligned}
\left\{ \bm{{\hat r ,\hat \theta}} \right\} = \mathop {\arg \min }\limits_{\bm{r, \theta}} \mathop {\min }\limits_{\bm {\Xi }_m} \left\| {{{\bm {\mathcal{V}}}_{sm}} - {{\bf{A}}_m}\left( \bm {r, \theta} \right){\bf{\Xi }}_m} \right\|_F^2.
\end{aligned}
\end{array}\
\end{equation}
Here, ${\bf{\Xi }}_m$ represents a nuisance parameter matrix associate with the signal covariance $\mathbb{E} \left[ {{{\bf{S}}_m}{\bf{S}}_m^H} \right]$, where $\mathbb{E} (\cdot) $ denotes the statistical expectation operator, and ${\left\|  \cdot  \right\|_F}$ indicates the Frobenius norm. The optimization objective is designed to minimize the subspace fitting error between the column space of the steering matrix ${{\bf{A}}_m}\left( \bm {r, \theta} \right)$ and the estimated signal subspace ${{\bm {\mathcal{V}}}_{sm}}$ in the least-squares sense. This formulation is justified by the fact that the steering vectors in ${{\bf{A}}_m}\left( \bm {r, \theta} \right)$ constitute an orthogonal basis for the signal subspace ${\bm{\mathcal{V}}}_{sm}$. For a given ${{\bf{A}}_m}\left( \bm {r, \theta} \right)$, the optimal nuisance parameter can be derived as ${\bf{\hat {\Xi} }} = {\bf{A}}_m^\dag \left( {r, \theta} \right){ {\bm{\mathcal{V}}}_{sm}}$, where ${\left(  \cdot  \right)^\dag }$ denotes the Moore-Penrose pseudo-inverse operation. Consequently, the maximum likelihood estimation (MLE) of the unknown parameters $\bm{\theta}$ and $\bm{r}$ across all subcarriers can be formulated as
\begin{equation}
\label{eq:eq-12}
\begin{array}{c}
\begin{aligned}
\left\{ \!\bm{{\hat r ,\hat \theta}}\! \right\}\! \!= \!\mathop {\arg \min }\limits_{\bm{r, \theta}}\!\! \!\sum\limits_{m = 0}^{M - 1}\!\!{ \left\| \left( {{\bf{I}} \!-\! {{\bf{A}}_m}\!\left( \bm {r, \theta} \right)\!{\bf{A}}_m^\dag \!\left( \bm {r, \theta} \right)} \right)\!{\bm{{\cal V}}_{sm}} \right\|_F^2},
\end{aligned}
\end{array}\
\vspace{-1.5mm}\end{equation}where $\bf {I}$ denotes identity matrix. By leveraging pseudo-inverse matrix properties and the trace-Frobenius norm relationship, we obtain~\cite{bohme1984estimation}
\vspace{-1mm}
\begin{equation}
\label{eq:eq-13}
\begin{array}{c}
\begin{aligned}
&\left\{ \bm{\hat r ,\hat \theta} \right\}\! = \!\mathop {\arg \max }\limits_{\bm{r, \theta}} \!\!\sum\limits_{m = 0}^{M - 1} \!\! { {\rm{tr}}\left( {{{{\bf{P}}_{{{\bf{A}}_m}}} \! \left( \bm {r, \theta} \right)}{{\bm {\mathcal{V}}}_{sm}}{{\bm {\mathcal{V}}}_{sm}^H}} \right)},
\end{aligned}
\end{array}\
\vspace{-0.5mm}\end{equation}
where projection matrix ${{{\bf{P}}_{{{\bf{A}}_m}}}\left( \bm {r, \theta} \right)} = {{\bf{A}}_m}{\left( {{\bf{A}}_m^H{{\bf{A}}_m}} \right)^{ - 1}}{\bf{A}}_m^H$ maps vectors onto the column space of ${{\bf{A}}_m}\left( \bm {r, \theta} \right)$, and $\rm{tr}( \cdot )$ denotes the matric trace operator. The complex structure of ${{\bf{A}}_m}\left( \bm {r, \theta} \right)$ induces a highly non-convex optimization landscape, characterized by multiple local maxima in proximity to the global optimum. To mitigate this computational challenge, we employ the single-target approximation method~\cite{pesavento2023three}, which reduces the search space dimensionality through the following formulation:
\begin{equation}
\label{eq:eq-14}
\left\{ \hat{r}_p, \hat{\theta}_p \right\}_{p=1}^P \!\!\!=\!
\underset{\substack{
    S \subseteq \mathcal{S} \\
    |S|=P \\
    (r,\theta)\in S
}}{\mathrm{arg\,max}}\!
\left[
    \sum_{m=0}^{M-1}\!
    \mathrm{tr}\!\left(
        \mathbf{P}_{\mathbf{a}_m}(r,\theta)
        \bm{\mathcal{V}}_{sm}
        \bm{\mathcal{V}}_{sm}^H
    \right)\!
\right]\!,
\end{equation}
where $\mathcal{S} = \{(r,\theta) \mid r_{\min} \leq r \leq r_{\max},\ \theta_{\min} \leq \theta \leq \theta_{\max}\}$ defines the parameter search space, and the constraint $|S|=P$ enforces selection of exactly $P$ solutions. This formulation enables efficient estimation of multiple targets while maintaining computational tractability, and
\begin{equation}
\label{eq:eq-15}
\begin{array}{c}
\begin{aligned}
{r_{n}} &= \sqrt {{{r}^2} + \delta _n^2{d^2} - 2r{\delta _n}d\cos \theta },\\
{{\bf{a}}_m}({r},{\theta}) &= {\left[ {\begin{array}{*{20}{c}}{{e^{ - j2\pi {f_m} \frac{r_1}{{{c}}} }},}&{ \ldots ,}&{{e^{ - j2\pi {f_m} \frac{r_N}{{{c}}} }}}\end{array}} \right]^T}.
\end{aligned}
\end{array}\
\vspace{-1mm}\end{equation}

It should be noted that the steering vector ${{\bf{a}}_m}({r},{\theta})$ exhibits dependence on both $\theta$ and $r$, as evident from the functional relationship of $r_n$ in Eq.~(\ref{eq:eq-15}). Moreover, the optimization objective can be equivalently formulated as either maximizing the cumulative projection $\sum\nolimits_{m = 0}^{M - 1} {{\rm{tr}}\left( {{{\bf {P}}_{{{\bf{a}}_m}}}\left( {r,\theta } \right){{\bm {\mathcal{V}}}_{sm}} {\bm {\mathcal{V}}}_{sm}^H} \right)}$, or equivalently, maximizing its complement\vspace{-1.5mm}
\begin{equation}
\label{eq:eq-16}
\begin{array}{c}
\begin{aligned}

1/\left( M - \sum\nolimits_{m = 0}^{M - 1} {{\rm{tr}}\left( {{{\bf {P}}_{{{\bf{a}}_m}}}\left( {r,\theta } \right){{\bm {\mathcal{V}}}_{sm}} {\bm {\mathcal{V}}}_{sm}^H} \right)}\right).

\end{aligned}
\end{array}\
\vspace{-1.5mm}\end{equation}
By substituting ${{{\bf{P}}_{{{\bf{a}}_m}}}\left( {r, \theta} \right)} = {{\bf{a}}_m}{\left( {{\bf{a}}_m^H{{\bf{a}}_m}} \right)^{ - 1}}{\bf{a}}_m^H$ and utilizing the fundamental properties of the trace operator, Eq.~(\ref{eq:eq-14}) can be simplified as
\begin{equation}
\label{eq:eq-17}
\begin{array}{c}
\begin{aligned}

\left\{ \hat{r}_p, \hat{\theta}_p \right\}_{p=1}^P & =
\underset{\substack{
    S \subseteq \mathcal{S} \\
    |S|=P \\
    (r,\theta)\in S
}}{\mathrm{arg\,max}}
 \frac{1}{{M - \sum\limits_{m = 0}^{M - 1} {\dfrac{{{{\left\| {{\bm {\mathcal{V}}}_{sm}^H{{\bf{a}}_m}(r,\theta )} \right\|}^2}}}{{{{\left\| {{{\bf{a}}_m}(r,\theta )} \right\|}^2}}}} }} \\

&\!\Leftrightarrow\! \underset{\substack{
    S \subseteq \mathcal{S} \\
    |S|=P \\
    (r,\theta)\in S
}}{\mathrm{arg\,max}}
 \underbrace{\!\!\left[ \!MN \!\!-\!\!\sum_{m=0}^{M-1} \!\! \|\bm{\mathcal{V}}_{sm}^H \mathbf{a}_m(r,\theta)\|^2 \right]^{-1}}_{\mathcal{J}(r,\theta)}.

\end{aligned}
\end{array}\
\vspace{-1mm}\end{equation}
By exploiting the orthogonality between the noise subspace ${{\bm {\mathcal{U}}}_{nm}}$ and signal subspace ${{\bm {\mathcal{V}}}_{sm}}$ based upon Eq.~(\ref{eq:eq-17}), we have
\vspace{-2mm}
\begin{equation}
\label{eq:eq-18}
\begin{array}{c}
\begin{aligned}

{\mathcal{J}(r,\theta)} & \!=\! \frac{1}{{\sum\limits_{m = 0}^{M - 1} {{{\left\| {{{\bf{a}}_m}(r,\theta )} \right\|}^2} \!-\! \sum\limits_{m = 0}^{M - 1} {{{\left\| {{\bm {\mathcal{V}}}_{sm}^H{{\bf{a}}_m}(r,\theta )} \right\|}^2}} } }} \\

&\!=\!  \frac{1}{{\sum\limits_{m = 0}^{M - 1} {{{\left\| {{\bm {\mathcal{U}}}_{nm}^H{{\bf{a}}_m}(r,\theta )} \right\|}^2}}}}.

\end{aligned}
\end{array}\
\end{equation}

\newcounter{TempEqCnt2}
\setcounter{TempEqCnt2}{\value{equation}}
\setcounter{equation}{21}
\begin{figure*}[hb]
\hrulefill
\begin{equation}
\label{eq:eq-23}
\begin{array}{c}
\begin{aligned}

 &{{\bf{\Sigma}}_m} \buildrel \Delta \over = \frac{1}{K} {{\bf{A}}_m}\left( \bm {r, \theta} \right){{\bf{S}}_m}{{\bf{S}}_m^H}{{\bf{A}}_m^H}\left( \bm {r, \theta} \right)  \\

 &\!\!\!=\! \frac{1}{K} \left[ {\begin{array}{*{20}{c}}
{{e^{ - j{k_m}{r_{1,1}}}}}& \ldots &{{e^{ - j{k_m}{r_{P,1}}}}}\\
 \vdots & \ddots & \vdots \\
{{e^{ - j{k_m}{r_{1,N}}}}}& \ldots &{{e^{ - j{k_m}{r_{P,N}}}}}
\end{array}} \right]\left[ {\begin{array}{*{20}{c}}
{{{\bf{\tilde s}}_{m, 1}^H}{\bf{\tilde s}}_{m, 1}}& \ldots &0\\
 \vdots & \ddots & \vdots \\
0& \ldots &{{{\bf{\tilde s}}_{m, P}^H}{\bf{\tilde s}}_{m, P}}
\end{array}} \right]\left[ {\begin{array}{*{20}{c}}
{{e^{j{k_m}{r_{1,N}}}}}& \ldots &{{e^{j{k_m}{r_{1,N}}}}}\\
 \vdots & \ddots & \vdots \\
{{e^{j{k_m}{r_{P,1}}}}}& \ldots &{{e^{ - j{k_m}{r_{P,N}}}}}
\end{array}} \right]\\

 &\!\!\!=\! \frac{1}{K} \!\!\!\left[\!\!\! {\begin{array}{*{20}{c}}
 {{{\bf{\tilde s}}_{m, 1}^H}{\bf{\tilde s}}_{m, 1} \!+\!  \ldots  \!+\! {{\bf{\tilde s}}_{m, P}^H}{\bf{\tilde s}}_{m, P}} & \ldots & {{{{e^{j{k_m}({r_{1,N}} - {r_{1,1}})}}}} {{\bf{\tilde s}}_{m, 1}^H}{\bf{\tilde s}}_{m, 1} \!+\!  \ldots  \!+\! {{{e^{j{k_m}({r_{P,N}} - {r_{P,1}})}}}}  {{\bf{\tilde s}}_{m, P}^H}{\bf{\tilde s}}_{m, P}} \\
 \vdots & \ddots & \vdots \\
 {{{{e^{ - j{k_m}({r_{1,N}} - {r_{1,1}})}}}} {{\bf{\tilde s}}_{m, 1}^H}{\bf{\tilde s}}_{m, 1} \!+\!  \ldots  \!+\! {{{e^{ - j{k_m}({r_{P,N}} - {r_{P,1}})}}}} {{\bf{\tilde s}}_{m, P}^H}{\bf{\tilde s}}_{m, P}} & \ldots &

 {{{\bf{\tilde s}}_{m, 1}^H}{\bf{\tilde s}}_{m, 1}  \!+\!  \ldots  \!+\! {{\bf{\tilde s}}_{m, P}^H}{\bf{\tilde s}}_{m, P}}
\end{array}} \!\!\!\right]\!\!\!,

\end{aligned}
\end{array}\
\end{equation}
\end{figure*}
\setcounter{equation}{\value{TempEqCnt2}}
\vspace{-1mm}
\subsection{The Proposed Low-Complexity MUSIC Algorithm}
Considering a ULA configuration with $N = 2 N' + 1$ elements, the array index ${\delta _n}$ defined in Section~\ref{sec:sig} can be reformulated as ${\delta _n} = n - 1 - N'$. To simplify Eq.~(\ref{eq:eq-1}), we employ the Fresnel approximation through second-order Taylor expansion~\cite{10934790}, expressed as
\begin{equation}
\label{eq:eq-22}
\begin{array}{c}
\begin{aligned}
{r_{p,n}} &\approx r_p\left( {1 + \dfrac{{{\delta _n^2}{d^2}}}{{2{r_p^2}}} - \dfrac{{{\delta _n}d}}{r_p}\cos \theta_p  - \dfrac{{{\delta _n^2}{d^2}{{\cos }^2}\theta_p }}{{2{r_p^2}}}} \right)\\
 &\approx r_p - {\delta _n}d\cos \theta_p + \dfrac{{{\delta _n^2}{d^2}}}{{2r_p}}{\sin ^2}\theta_p.\\
\end{aligned}
\end{array}\
\end{equation}
Therefore, the phase component of the array steering vector ${{\bf{a}}_m}\left( {r_p,\theta_p } \right)$ can be approximated as\vspace{-0.5mm}
\begin{equation}
\label{eq:eq-19}
\begin{array}{c}
\begin{aligned}
\begin{array}{c}
- 2\pi {f_m}\dfrac{{{r_{p,n}}}}{c} \approx {\varphi _m} + {\gamma  _m}{\delta _n} + {\eta _m}{\delta _n^2},
\end{array}
\end{aligned}
\end{array}\
\vspace{-2.5mm}\end{equation}
where
\begin{equation}
\label{eq:eq-20}
\begin{array}{c}
\begin{aligned}
{\varphi _{m,p}} \!=\!  - \dfrac{{2\pi }}{{{\lambda _m}}}r_p, {\gamma _{m,p}} = \dfrac{{2\pi d}}{{{\lambda _m}}}\cos \theta_p, {\eta _{m,p}} =  - \dfrac{{\pi {d^2}}}{{{\lambda _m}r_p}}{\sin ^2}\theta_p.
\end{aligned}
\end{array}\
\end{equation}
Based on the preceding assumptions and the Fresnel approximation, the array steering vector ${{\bf{a}}_m}\left( {r,\theta } \right)$ simplifies to the following compact form:
\begin{equation}
\label{eq:eq-21}
{{\bf{a}}_m}(r,\theta ) = {e^{j{\varphi _m}}}\left[ {{e^{j\left( {{\gamma _m}n + {\eta _m}{n^2}} \right)}}} \right],\quad n \in \{  - N', \ldots ,N'\}.
\end{equation}

Building upon Eq.~(\ref{eq:eq-22}), we reconsider the covariance matrix from Eq.~(\ref{eq:eq-8}) by temporarily neglecting the path loss coefficients $\bm \beta_m$ and noise ${{\bf{W}}_m}$ for analytical simplicity. The resulting covariance matrix is calculated as Eq.~(\ref{eq:eq-23}) shown at the bottom of this page, where $k_m = 2 \pi / \lambda_m$ is the wave number, and ${{\bf{\tilde s}}_{m, p}} = \left[ {\begin{array}{*{20}{c}} {{s_{0, m, p}},}&{ \ldots ,}&{{s_{K-1, m, p}}} \end{array}} \right]^T \in {{\mathbb C}^{K \times 1}}$. Assuming the transmitted signals from different targets are independent of each other, we have
\setcounter{equation}{22}
\begin{equation}
\label{eq:eq-24}
\begin{array}{c}
\begin{aligned}
{{\bf{\tilde s}}_{m, p}^H}{\bf{\tilde s}}_{m, q} = \left\{ {\begin{array}{*{20}{c}}
{\sum\limits_{k = 0}^{K-1} {{s_{k, m, p}^*}s_{k, m, q},p = q} },\\
{0,p \ne q}.
\end{array}} \right.
\end{aligned}
\end{array}\
\vspace{-1mm}\end{equation}
Additionally, by examining the phase characteristics of the anti-diagonal elements in the covariance matrix from Eq.~(\ref{eq:eq-23}), we have
\begin{equation}
\label{eq:eq-25}
\begin{array}{c}
\begin{aligned}

 - j{k_m}({r_{p,N}} - {r_{p,1}}) = 2{k_m}d\cos {\theta _p}N'.

\end{aligned}
\end{array}\
\end{equation}
This reveals that the anti-diagonal elements of the covariance matrix are exclusively determined by the angular-dependent parameters ${\gamma _{m,p}}$, which are functions of the angles $\theta_{p}$.

Then, by substituting Eq.~(\ref{eq:eq-25}) into the anti-diagonal entries of the covariance matrix in Eq.~(\ref{eq:eq-23}) and organizing them into a vector form, we obtain\vspace{-1mm}
\begin{equation}
\label{eq:eq-26}
\mathbf{\bar{y}}_m =
\sum\limits_{p=1}^{P} \chi_{m,p}
{\bm{\gamma}}_{m,p},
\vspace{-1.5mm}\end{equation}
where ${\bm{\gamma}}_{m,p} = [e^{-j 2N' \gamma_{m,p}}, \dots, e^{j 2N' \gamma_{m,p}}]^T \in {{\mathbb C}^{(2N' + 1) \times 1}}$, and ${\chi _{m, p}} = \frac{1}{K}{{\bf{\tilde s}}_{m, p}^H}{\bf{\tilde s}}_{m, p}$ denotes the normalized signal energy. Notably, the elements of ${{\bf{\bar y}}_{m}}$ are independent of the angular parameter $\theta_p$, allowing the separation of the distance $r_p$ and angle $\theta_p$ estimation in Eq.~(\ref{eq:eq-1}). Nevertheless, the vector ${{\bf{\bar y}}_{m}}$ is derived from a single snapshot, which may introduce signal correlation among multiple targets. To address this issue, we implement spatial smoothing to construct an enhanced covariance matrix with rank exceeding $P$. This technique involves partitioning ${{\bf{\bar y}}_{m}}$ into $L$  overlapping subvectors, each of dimension $2N' + 2 -L$. Both $2N' + 2 -L$ and $L$ should be greater than the number of estimated targets $P$, yielding
\begin{equation}
\label{eq:eq-27}
\mathbf{\tilde{y}}_m(l) =
\sum\limits_{p = 1}^{P} \chi_{m, p} {\tilde{\bm{\gamma}}}_{m,p}
= \mathbf{\tilde{A}}(\gamma_m) \mathbf{\tilde{s}}_m(l),
\end{equation}
where
\begin{equation}
\label{eq:eq-28}
\begin{array}{c}
\begin{aligned}

\tilde{\bm{\gamma}}_{m,p} &\!=\! \left[ e^{-j2(N'+l-j)\gamma_{m,p}} \right]_{j=1}^{L}, {\bf{\tilde A}}\left( {{\gamma _{m}}} \right) \!=\! \left[ {{\bf{\tilde a}}\left( {{\gamma _{m,p}}} \right)} \right]_{p=1}^P,\\

{\bf{\tilde a}}\left( {{\gamma _{m,p}}} \right) &\!=\! {\left[ {\begin{array}{*{20}{c}}
{{e^{ - j 2  i {\gamma _{m,p}}}}}
\end{array}} \right]_{i = N' + 1}^{L - N'}}, \mathbf{\tilde{s}}_m(l) \!=\! \left[ \chi_{m,p} e^{jl\gamma_{m,p}} \right]_{p=1}^P.\\

\end{aligned}
\end{array}\
\end{equation}
It is noteworthy that a structural similarity exists between Eqs.~(\ref{eq:eq-7}) and (\ref{eq:eq-27}), where ${\bf{\tilde A}}\left( {{\gamma _{m}}} \right)$ resembles a conventional array steering matrix. The spatial smoothing covariance matrix, constructed from the $L$ overlapping subvectors for estimating the $P$ angles encoded in $\gamma _{m,P}$, is computed as
\begin{equation}
\label{eq:eq-29}
\begin{array}{c}
\begin{aligned}

{{{\bf{\tilde \Sigma}}}_m} = \frac{1}{L}\sum\limits_{l = 1}^L {{{{\bf{\tilde y}}}_m}\left( l \right){\bf{\tilde y}}{{_m^H}}\left( l \right)}.

\end{aligned}
\end{array}\
\end{equation}

By leveraging the Fresnel approximation, which decouples the angle information in Eq.~(\ref{eq:eq-24}),  angle estimation is initially performed using the spatially smoothed covariance matrix of Eq.~(\ref{eq:eq-29}). Subsequently, the corresponding distance parameter is obtained through a computationally efficient one-dimensional (1D) spectral search based on the estimated angle parameters, thereby effectively reducing the complexity of the joint estimation problem.
\vspace{-1.5mm}
\section{Numerical Results}
\label{sec:result}

This section presents numerical results to validate the effectiveness of the proposed subspace fitting-based algorithm and its modified MUSIC method. Unless otherwise specified, the system parameters are configured as follows: $f_c = 28$ GHz, $N = 128$, $K = 200$, $M = 64$, $\Delta f = 480$ kHz, and $L = 50$. Moreover, the simulation scenario considers $P = 2$, whose angles and distances are randomly generated within the near-field region. The NMSEs are evaluated through extensive Monte Carlo simulations with $200$ independent trials to ensure statistical reliability.

\begin{figure}
\begin{center}
\subfloat[NMSEs versus SNRs under both narrowband and wideband conditions.] {
        \label{fig:fig-3a}
        \centering
		\includegraphics[width=0.46\linewidth]{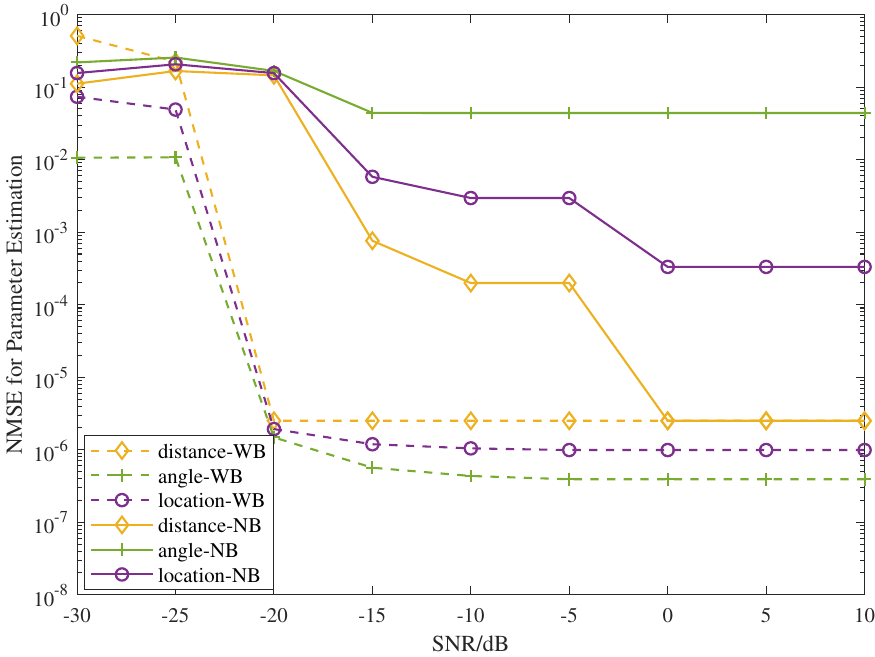}}
\hfil
 \subfloat[NMSEs versus Bandwidth under a fixed number of subcarriers.]{
         \label{fig:fig-3b}
         \centering
         \includegraphics[width=0.46\linewidth]{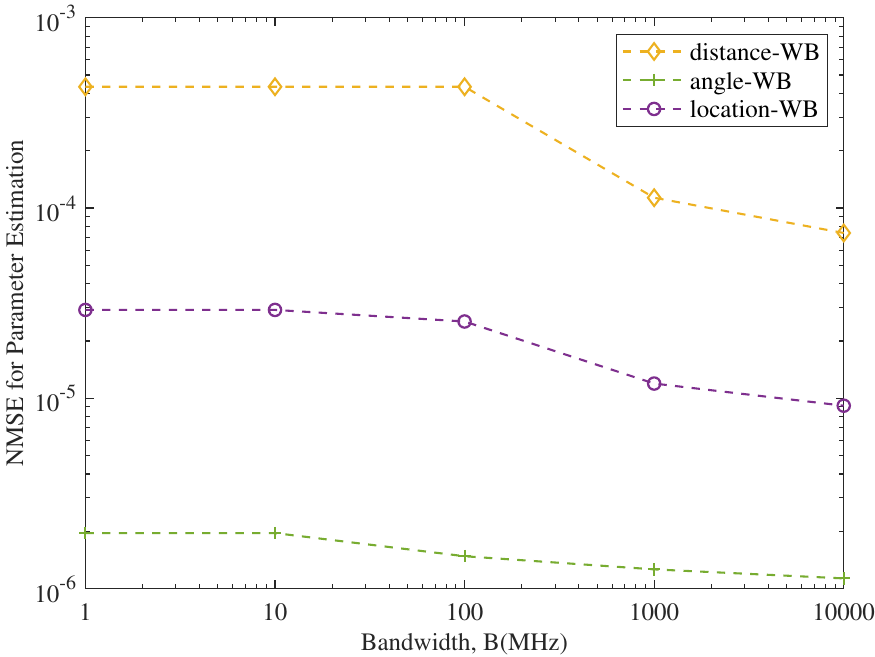}}
\end{center}\vspace{-2.5mm}
\caption{The performance of the proposed low-complexity MUSIC algorithm.}
\label{fig:fig-3}
\end{figure}

Figs.~\ref{fig:fig-2} and~\ref{fig:fig-3} present a performance evaluation of the normalised mean square errors (NMSEs) for both the subspace fitting-based MUSIC method and the proposed low-complexity MUSIC algorithm. The analysis spans a wide range of form $-30$dB to $10$dB and bandwidths form $1$ MHz to $10^6$ MHz. The legend notation follows the convention "X-Y", where
\begin{itemize}
\item "location-NB" indicates 2D (distance and angle) parameter estimation under narrowband conditions.
\item "distance-WB" represents distance estimation performance in wideband scenarios.
\end{itemize}

\vspace{-1mm}
\subsection{Performance of Subspace Fitting-based MUSIC Method}
\begin{figure}
\begin{center}
\subfloat[][NMSEs versus SNRs under both narrowband and wideband conditions.] {
        \label{fig:fig-2a}
        \centering
		\includegraphics[width=0.46\linewidth]{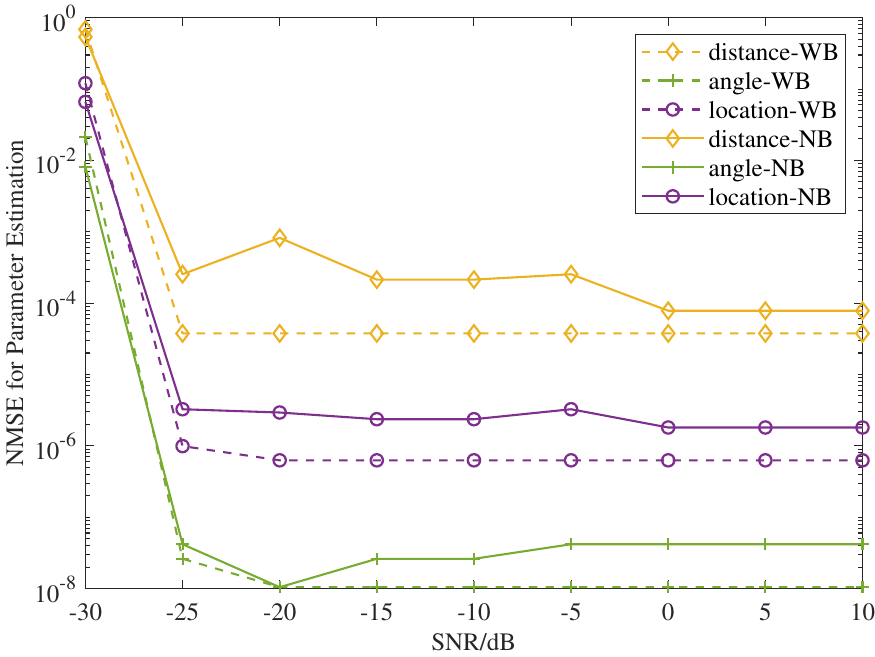}}
\hfil
 \subfloat[][NMSEs versus Bandwidth under a fixed number of subcarriers.]{
         \label{fig:fig-2b}
         \centering
         \includegraphics[width=0.46\linewidth]{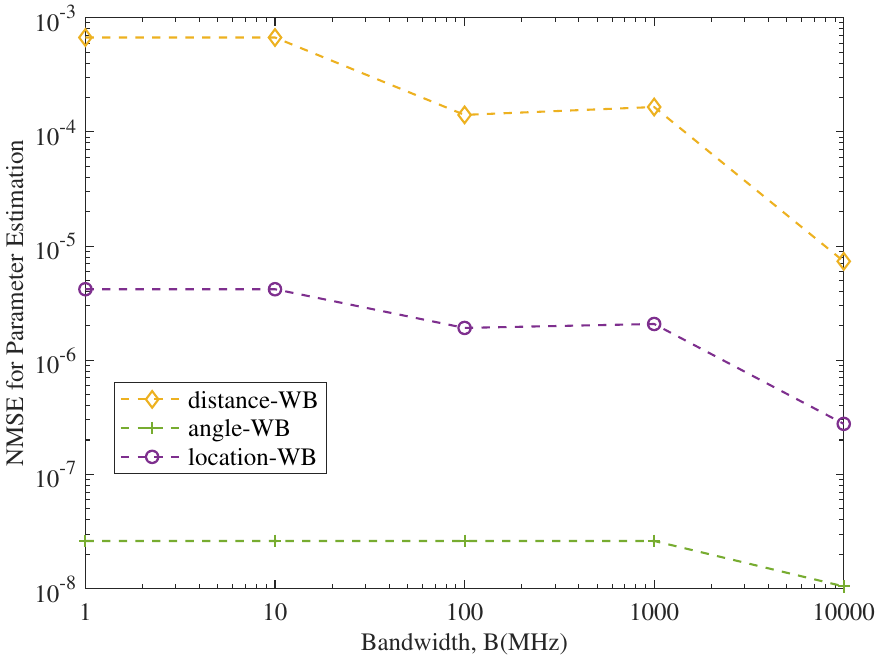}}
\end{center}\vspace{-2mm}
\caption{The performance of subspace fitting-based MUSIC method.}
\label{fig:fig-2}
\end{figure}

Fig.~\ref{fig:fig-2a} demonstrates the NMSE performance of the subspace fitting-based MUSIC algorithm under varying SNR conditions for both narrowband and wideband configurations. The number of subcarriers is fixed at $M = 64$ while maintaining distinct subcarrier spacing: $480$ Hz for narrowband and $480 \times 10^5$ Hz for wideband. It can be observed that the wideband implementation consistently outperforms its narrowband counterpart in both distance and angle estimation tasks across the entire SNR range even in low SNR regimes (eg., below $-15$dB). Additionally, the performance of 2D parameter estimation, as expected, lies between the individual distance and angle estimation curves, reflecting the combined influence of the two estimation processes.

Fig.~\ref{fig:fig-2b} investigates the impact of varying bandwidth $B$ under the condition of a fixed number of subcarriers ($M = 64$) and adjusting the subcarrier spacing $\Delta f$ accordingly. It can be observed that increasing bandwidth generally improves sensing performance, with particularly significant gains in distance estimation accuracy.
\vspace{-2mm}

\subsection{Performance of Low-Complexity MUSIC Algorithm}
Similarly, Fig.~\ref{fig:fig-3a} illustrates the effect of SNRs on NMSEs for the proposed low-complexity MUSIC algorithm. The experimental setup maintains identical parameters to those used in the subspace fitting-based MUSIC method for comparative analysis. The experimental results demonstrate that the wideband complexity implementation significantly outperforms the narrowband configuration
across all SNR conditions. Substantial performance gains are achieved in both distance and angle estimation. At an SNR of $-15$ dB, the wideband implementation achieves an NMSE of $2.51 \times 10^{-6}$, representing a significant improvement in distance estimation accuracy compared to the narrowband NMSE of $7.61 \times 10^{-4}$.

Similar to the observation in Fig.~\ref{fig:fig-2a}, bandwidth expansion generally enhances estimation performance in Fig.~\ref{fig:fig-3a}. However, the NMSE performance in Figs.~\ref{fig:fig-3a} and~\ref{fig:fig-3b} are slightly degraded compared to Fig.~\ref{fig:fig-2a} and~\ref{fig:fig-2b}, respectively, primarily due to the Taylor approximation employed in the low-complexity MUSIC algorithm, which introduces minor sensing accuracy loss while maintaining computational efficiency.

\section{Conclusion}
This paper has proposed a subspace fitting-based MUSIC method for wideband near-field localization in multi-target OFDM systems. A frequency-domain near-field signal model has been first established to enable simultaneous distance and angle estimation. To reduce complexity, the Fresnel approximation-based variant has been introduced, effectively lowering the algorithm’s search dimensionality. Numerical results have demonstrated the enhanced parameter resolution and computational efficiency of the proposed approach. The framework provides a practical solution for high-precision near-field positioning in next-generation networks, with potential extensions to hybrid field scenarios.

\bibliographystyle{IEEEtran}
\bibliography{reference}



\end{document}